\documentclass[a4paper,fontsize=12pt]{scrartcl}

\usepackage[a4paper, total={6.5in, 7.5in}]{geometry}

\usepackage{xspace}
\usepackage{epsfig}
\usepackage{xcolor}
\usepackage{syntax}
\usepackage[fleqn]{amsmath}
\usepackage{amssymb}
\usepackage{semantic}
\usepackage{hyperref}
\usepackage{lmodern}
\usepackage[T1]{fontenc}
\usepackage{makeidx}
\usepackage{makecell}
\usepackage{float}
\usepackage{alltt}

\makeindex
\title{\LARGE{QRtree - Decision Tree dialect specification of QRscript}\\
{\Large (Version 0.9)}}
\author{}

\begin{document}
\maketitle

\begin{abstract}
This document \cite{QRtree} reports the specifications of QRtree, a specific dialect of QRscript that can be used for the inclusion of decision trees (or other similar high-level languages) within a QR code. A QR code containing an executable code is called an executable QR code (eQR code).
QRscript is a series of rules on how to embed a programming language into a QR code in order to obtain an eQR code.

Authors of this specification document are:
\begin{itemize}
    \item Stefano Scanzio (CNR-IEIIT, \href{https://www.skenz.it/ss}{https://www.skenz.it/ss}, stefano.scanzio[at]cnr.it)
    \item Matteo Rosani (matteo.rosani[at]gmail.com)
    \item Mattia Scamuzzi (mattia.scamuzzi[at]gmail.com)
    \item Gianluca Cena (CNR-IEIIT, gianluca.cena[at]cnr.it)
\end{itemize}
\end{abstract}

\clearpage
\tableofcontents

\newpage

\pagestyle{myheadings}
\markright{QRtree - Decision Tree dialect specification of QRscript (Version 0.9)}

\section{Scope}
This specification document specifies the syntax and semantics of QRtree, which is a specific \textit{dialect} of QRscript particularly suited to represent decision trees without chance nodes. The term ``dialect'' identifies one of the possible sub-languages that can be encoded inside of an eQR code via QRscript.

This specification will describe an \textit{intermediate representation} of QRtree, made through a language derived by the three-address code. It will then define the transformation rules from the intermediate representation to a \textit{binary code}. The latter is a binary representation called eQRtreebytecode. These rules can also be applied inversely to transform the eQRtreeBytecode into the intermediate representation.

This specification document will pay particular attention to the creation of a compact eQRtreebytecode, as the maximum number of bits that can be stored in a QR code is, at the time of writing, equal to 2953 bytes (in the case of QR code version 40 with a ``low'' error correction level).

\section{Conformance}
A completely conforming implementation of QRscript must provide and support all the functionalities mandated in this specification. The word \textit{must} highlights these mandatory specifications. Some specific names that are important in this specification were highlighted in \textit{italic}.

\section{Normative References}
The following referenced documents are indispensable for the application of this document.
For dated references, only the edition cited applies. For undated references, the latest edition of the referenced document (including any amendments) applies.

\begin{enumerate}
\item The QR code standard, ISO/IEC 18004:2015 \cite{QRstandard}
\item The UTF-8 standard, RFC 3629 \cite{rfc3629}
\item The ISO 7-bit coded character, ISO/IEC 646, \cite{ISO646-1991}
\item Floating-point arithmetic standard, IEEE Std 754, \cite{IEEE754-2019}
\end{enumerate}

\section{Overview}
This document specifies how to include a program that complies with the QRtree dialect into an eQR code.

The possibility to insert a programming language inside a QR code and, particularly, the basis of the QRtree dialect presented in this specification were first and preliminarily presented in \cite{9921530}.

\begin{figure}[ht]
	\begin{center}
	\includegraphics[width=0.8\columnwidth]{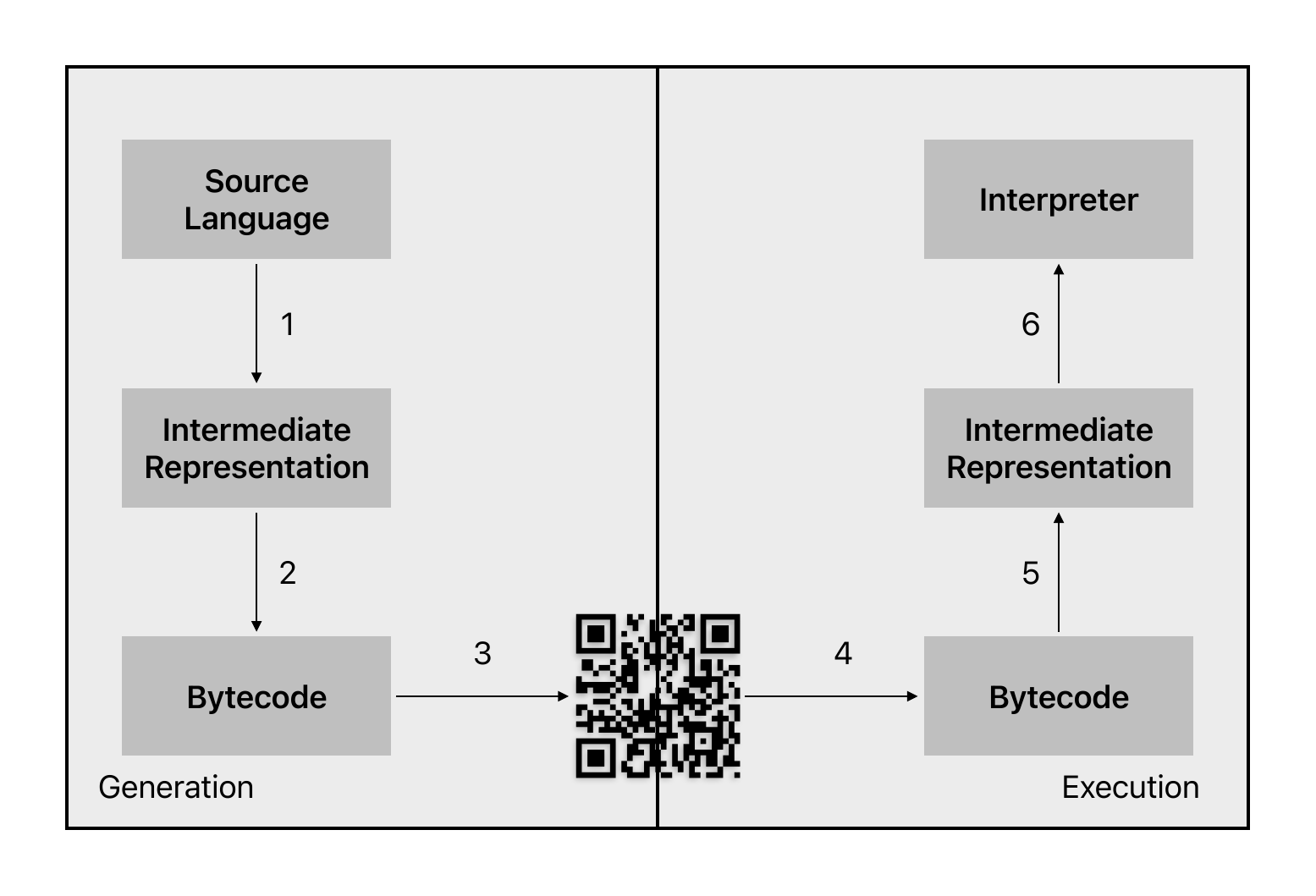}
	\end{center}
	\caption{Chain of usage of eQR code technology (generation and execution phases)}
	\label{fig:generationAndExecution}
\end{figure}

The whole utilization chain of this technology can be schematized as shown in Fig.~\ref{fig:generationAndExecution} in which the process is divided into a \textit{generation} phase and an \textit{execution} phase.

The generation phase leads to the creation of the eQR code, and it is detailed below:
\begin{enumerate}
    \item A \textit{high-level language} (Source Language) is translated into an \textit{intermediate representation} (Intermediate Representation). It does not matter which high-level language is used as it is not the topic of this specification document. Different high-level languages can be translated to the same intermediate representation.
    \item The intermediate representation is translated to a \textit{binary representation} (Bytecode) named \textit{eQRbytecode}.
    \item The eQRbytecode is transformed into an \textit{eQR code} containing the bytecode.
\end{enumerate}

The execution phase leads to the reading of the eQR code and its usage, and it is detailed below:
\begin{enumerate}
    \setcounter{enumi}{3}
    \item The eQR code is read and transformed in eQRbytecode.
    \item The eQRbytecode is converted into an intermediate representation
    \item The intermediate representation is executed inside of a \textit{virtual machine} on board the \textit{end-user device}. An end-user device is any apparatus capable of executing the virtual machine (i.e., a smartphone, a PC, an embedded device, etc.)
\end{enumerate}

The QRtree specification concerns the definition of the binary code that must be saved within the eQR code, and in particular, the part related to the QRtree dialect of QRscript. These specifications correspond to arrows 2 and 5 of Fig.~\ref{fig:generationAndExecution}, with the exclusion of the QRtree header, which is detailed in the relevant specification document \cite{QRscript-spec}.
In addition, the same specification document details the main characteristics regarding arrows 3 and 4 of Fig.~\ref{fig:generationAndExecution}. Arrows 1 and 6 are not subject to the specification but have been shown in Fig.~\ref{fig:generationAndExecution} to give a general idea of the applicability of the technique.

In particular, among the parts described in this QRtree specification document, there is an intermediate (low-level) language, similar to three-address code, which allows the coding, in a simplified way, of decision trees without chance nodes. Chance nodes are particular nodes of a decision tree that permit the coding of probabilistic decisions. QRtree is not limited to the representation of decision trees, but more generally, all the possible \textit{high-level languages} that can be transformed into the intermediate representation are possible candidates.

This specification defines and details all the commands available in the intermediate representation, the rules for converting them into the QRtreebytecode binary code, and the format that must be used to form the QRtreebytecode.

The following Section~\ref{sec:QRscript} will describe the headers definable in the QRbytecode, which, among other things, identify the dialect and allow reading more than one eQR code. This last possibility was inserted to try to mitigate the limitation of the amount of data that can be stored within an eQR code. Section~\ref{sec:QRtree} will define all the specific details of QRtree.

\section{QRScript and Dialects}
\label{sec:QRscript}
The first part of the QRbytecode containing the QRtree dialect follows the specifications defined in the specification document of QRscript \cite{QRscript-spec}, and it is named \textit{QRscript header}. It encodes five aspects. The first is related to the \textit{padding} that is used to make the length of the QRbytecode multiple of 8 bits, which is needed if the \textit{binary input mode} is used for the generation of the eQR code. The second is the \textit{continuation}, which allows the reading of more than one eQR code in order to try to mitigate the limitation of the amount of data that can be stored within an eQR code. The third is related to \textit{security} by defining a \textit{security profile} that identifies which security mechanisms are used to protect the eQR code data. The fourth is related to a \textit{URL} to be accessed in order to enhance the functionalities of the eQR code by executing a remote resource if a connection is available. The fifth and last aspect is the identification of the dialect (QRtree in the case of this specification document).

In particular, a QRbytecode starts with a \texttt{1} in the case the padding is not present, with the sequence \texttt{01} for 1 bit of padding, with the sequence \texttt{001} for 2 bit of padding, with the sequence \texttt{0001} for 3 bit of padding, and so on.
Then, the next bit is \texttt{0} in case all the QRbytecode can be stored within a single eQR code. Alternatively, the bit is equal to \texttt{1}, and the next four bits represent a \textit{sequence number} assigned to a specific eQR code, and the other four bits represent the \textit{sequence length}. These two groups of $4$ bits are extensible following the exponential encoding described in Appendix~\ref{app:C}. During the reconstruction of the QRbytecode, the application that performs the reading can use the sequence numbers to concatenate the binary code contained in the various eQR codes in the correct order.

After the bits related to the continuation, there are $4$ bits that identify the \textit{security profile} with possibly some extra information like a digital signature. The security profile number $0000$ corresponds to no security. This group of $4$ bits is extensible following the exponential encoding described in Appendix~\ref{app:C}.

After the bits related to security, there is $1$ bit to signal the presence of a \textit{URL}. If the bit is set to $1$, the following bytes identify a URL string encoded in UTF-8 \cite{rfc3629} using the end-of-text (EXT) character at the end of the string, corresponding to the sequence of bits \texttt{00000011}; otherwise, if the bit is set to $0$ the URL is not present.

After the bits related to the URL, there are $4$ bits that identify the dialect. This group of $4$ bits is extensible following the exponential encoding described in Appendix~\ref{app:C}. In particular, for the QRtree dialect, these bits are set to \texttt{0000}.

\section{QRtree dialect}
\label{sec:QRtree}
\begin{figure}
    \begin{center}
	\includegraphics[width=0.8\columnwidth]{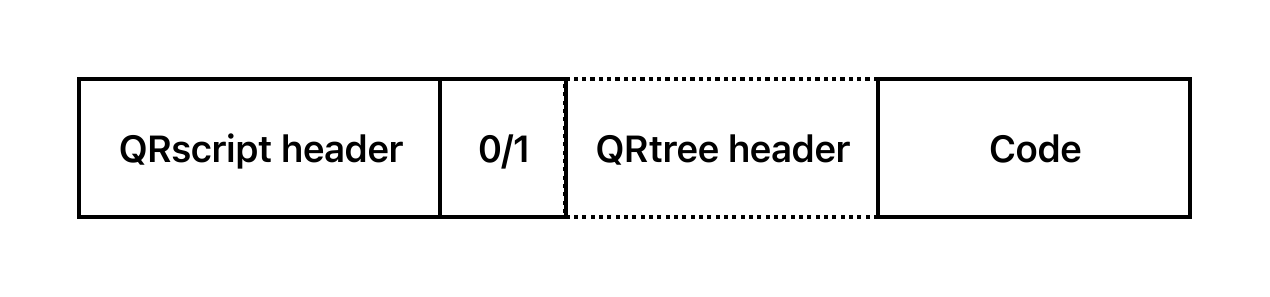}
	\end{center}
    \caption{Binary format of the eQR code (i.e., QRbytecode)}
    \label{fig:eQRcodeBinaryFormat}
\end{figure}

The QRtreebytecode can be divided into two sections: \textit{QRtree header}, which is optional and is described in Subsection~\ref{sec:header}, and \textit{code}, which is described in Subsection~\ref{sec:code}.
The implementation of the header and of the \textit{dictionary}, which will be explained next, are optional. Although their implementation is highly recommended, a compliant implementation of this specification document may not include these two features.
Fig.~\ref{fig:eQRcodeBinaryFormat} outlines the structure of how binary code is stored within an eQR code. In particular, after the \textit{QRscript header}, there is a bit that, if set to \texttt{1}, identifies the presence of a \textit{QRtree header}, which is optional. The QRtree header is then followed by the actual \textit{code}, which contains the instructions of the program that is coded into the eQR code.

Before the description of the \textit{QRtree header} and the \textit{code}, the next Section~\ref{sec:types} describes the data types available in the QRtree dialect, and some indications on how they are coded in the QRtreeBytecode.

\subsection{Data types}
\label{sec:types}
The constant data formats (literals) that can be used in QRtree are \textit{strings}, \textit{integers} numbers, and \textit{real} numbers.

\subsubsection{String: ASCII-7, UTF-8 and DICT}
\label{sub:string}
The encoding of the string type is very important because this type of data is widely used in decision trees, but it also occupies a lot of space; for this reason, three different encodings have been proposed, in which the first $2$ bits identify the type of encoding used: \texttt{00} (ASCII-7), \texttt{01} (UTF-8), \texttt{10} (DICT), while \texttt{11} has been left free for future use.

\begin{itemize}
\item \textbf{ASCII-7}\\ The ASCII-7 encoding \cite{ISO646-1991} is identified by the bit sequence \texttt{00}, followed by the sequence of characters that make up the string, and terminated by an end-of-string character. The chosen end-of-string character is \textit{end-of-text (EXT)} composed of the following bits \texttt{0000011}. The ASCII-7 encoding allows encoding a character in only 7 bits, but obviously, the kind of representable characters is limited. For example, the encoding of accented characters is not supported.

\item \textbf{UTF-8}\\ The UTF-8 encoding \cite{rfc3629} is identified by the bit sequence \texttt{01} followed by the sequence of characters that make up the string, and terminated by an end-of-string character. The chosen end-of-string character is \textit{end-of-text (EXT)} composed of the following bits \texttt{00000011}. The UTF-8 encoding allows encoding a much wider character set than ASCII-7. Furthermore, the encoding of characters with a minimum of 8 bits allows the encoding of other characters, for example, ideographic codes, with a greater number of bits (up to 4 bytes per character).

\item \textbf{DICT}\\ The DICT type is intended for the use of predefined strings in order to save space. Rather than memorizing the string inside the eQR code, the string is identified with a reference that occupies far less space. In particular, the following three types of DICT strings are identified: \textit{global}, \textit{specific}, and \textit{local}. The \textit{global} type strings are defined outside eQR codes because they are not specific for the application, but they are so general that they are effective for a multitude of applications (e.g., ``yes'', ``no'', etc.). Also, the \textit{specific} type strings are defined outside eQR codes, but they are specific for an application. The selection of the \textit{specific} dictionary to be used is done in the \textit{QRtree header} via the \texttt{DICT_SPEC_TYPE} command (see Subsection~\ref{sub:DICTSPECTYPE} for details). Both \textit{global} and \textit{specific} dictionaries are defined using JSON, following the specifications described in Appendix~\ref{app:A}. In particular, URLs containing the \textit{global} dictionary and potentially one or more \textit{specific} dictionaries in JSON format are saved inside the application.
In Appendix~\ref{app:A}, a simple example of a \textit{global} dictionary is reported. In case multiple \textit{specific} dictionaries are present, the \texttt{DICT_SPEC_TYPE} command specifies the number of the dictionary to be used based on the loading order (e.g., number 0 for the first \textit{specific} dictionary loaded, number 1 for the second \textit{specific} dictionary loaded, etc.). It is possible to load more \textit{specific} dictionaries using the \texttt{DICT_SPEC_TYPE} command multiple times in the \textit{QRtree header}; in this case, inside the eQR code, it is possible to use multiple \textit{specific} dictionaries, and a particular notation could be used for the DICT type. In order to select the right \textit{specific} dictionary to be used between the loaded ones, see further on in this paragraph. With regard to the \textit{local} type dictionary, it is defined inside the \textit{QRtree header} using the \texttt{DICT_LOCAL} command (see Subsection~\ref{sub:DICTLOCAL} for details). As strings occupy a lot of space, attention is needed in defining as local only the ones that appear more frequently in the eQR code. The definition of the \textit{local} dictionary or a part of it can be automated through a tool that automatically analyses strings appearing in the eQR code, and places in the \textit{local} dictionary more frequent ones. In case dictionaries contain multiple languages, the application will select the dictionary for the language with which the application has been configured. If the language is not present in the dictionary, the application will select the default dictionary, for example, the one for the English language. In this case, the application is in charge of setting the default dictionary. If the default dictionary is not set, the application selects the first that has been defined for the specific type that is selected (i.e., \textit{global}, \textit{specific}, or \textit{local}).

\begin{table}[H]
  \caption{Bits used to identify the type of dictionary (\textit{global}, \textit{local}, \textit{specific}). The symbol \texttt{-} indicates that this type of dictionary cannot be used, while the symbol \texttt{*} indicates that the dictionary can be used without writing any bits because it is the only dictionary type configured for that specific eQR code. The default configuration is \texttt{DICT_GLOBAL}, \texttt{DICT_SPEC} and \texttt{NO_DICT_LOCAL}}
  \label{tab:dictionary}

  \begin{center}
    \tabcolsep=0.18cm
    \def\arraystretch{1.28}
    \begin{tabular}{c|ccc}
    Dictionary & \textit{global} & \textit{specific} & \textit{local} \\
    Options & & & \\
    \hline    
    \makecell{\texttt{DICT_GLOBAL}\\\texttt{DICT_SPEC}\\\texttt{DICT_LOCAL}} & 00 & 01 & 1 \\
    \hline
    \makecell{\texttt{NO_DICT_GLOBAL}\\\texttt{NO_DICT_SPEC}\\\texttt{NO_DICT_LOCAL}} & - & - & - \\
    \hline
    \makecell{\texttt{DICT_GLOBAL}\\\texttt{NO_DICT_SPEC}\\\texttt{DICT_LOCAL}} & 0 & - & 1 \\
    \hline
    \makecell{\texttt{NO_DICT_GLOBAL}\\\texttt{DICT_SPEC}\\\texttt{DICT_LOCAL}} & - & 0 & 1 \\
    \hline
    \makecell{\texttt{DICT_GLOBAL}\\\texttt{DICT_SPEC}\\\texttt{NO_DICT_LOCAL}} & 0 & 1 & - \\
    \hline
    \makecell{\texttt{DICT_GLOBAL}\\\texttt{NO_DICT_SPEC}\\\texttt{NO_DICT_LOCAL}} & * & - & - \\
    \hline
    \makecell{\texttt{NO_DICT_GLOBAL}\\\texttt{DICT_SPEC}\\\texttt{NO_DICT_LOCAL}} & - & * & - \\
    \hline
    \makecell{\texttt{NO_DICT_GLOBAL}\\\texttt{NO_DICT_SPEC}\\\texttt{DICT_LOCAL}} & - & - & * \\
    \hline
   \end{tabular}
  \end{center}
\end{table}

The DICT type is identified by the sequence of bits \texttt{10}. The next bits, named \textit{dictionary\_type}, refer to the type of dictionary (\textit{global}, \textit{specific} or \textit{local}), and the number and type of bits to use depends on the configuration of some constants in the QRtree header, as shown in Table~\ref{tab:dictionary}. In particular, the default configuration, which is \texttt{DICT_GLOBAL}, \texttt{DICT_SPEC} and \texttt{NO_DICT_LOCAL}, identifies the presence of a \textit{global} and a \textit{specific} dictionary, and the absence of a \textit{local} dictionary.
If a \textit{local} dictionary is defined using the \texttt{DICT_LOCAL} command (Subsection~\ref{sub:DICTLOCAL}), the \texttt{DICT_LOCAL} constant is activated.
Instead, the command \texttt{DICT_TYPES} (Subsection~\ref{sub:DICTTYPES}) can be used to disable the \textit{global} dictionary by setting the \texttt{NO_DICT_GLOBAL} constant or the \textit{specific} dictionary, by setting the \texttt{NO_DICT_SPEC} constant.

Finally, the last bits of the DICT type represent the word contained inside the dictionary that was selected using the previous bits (i.e., \textit{global}, \textit{specific}, or \textit{local}). The word is selected using its position inside the dictionary, with a number of bits equal to the minimum number of bits necessary to encode all the words inside the selected dictionary. For example, if the dictionary contains 23 words, the words will be encoded and referred to using 5 bits. The number $0$ corresponds to the first word. In case multiple dictionaries of type \textit{specific} are present at the same time, they are concatenated in the order in which the \texttt{DICT_SPEC_TYPE} commands appear and, as in the previous case, the word is selected using its position within the \textit{specific} dictionaries concatenation, using a number of bits equal to the smallest number of bits used to encode all the words within the concatenation of dictionaries of type \textit{specific}.

\begin{figure}
    \begin{center}
	\includegraphics[width=0.35\columnwidth]{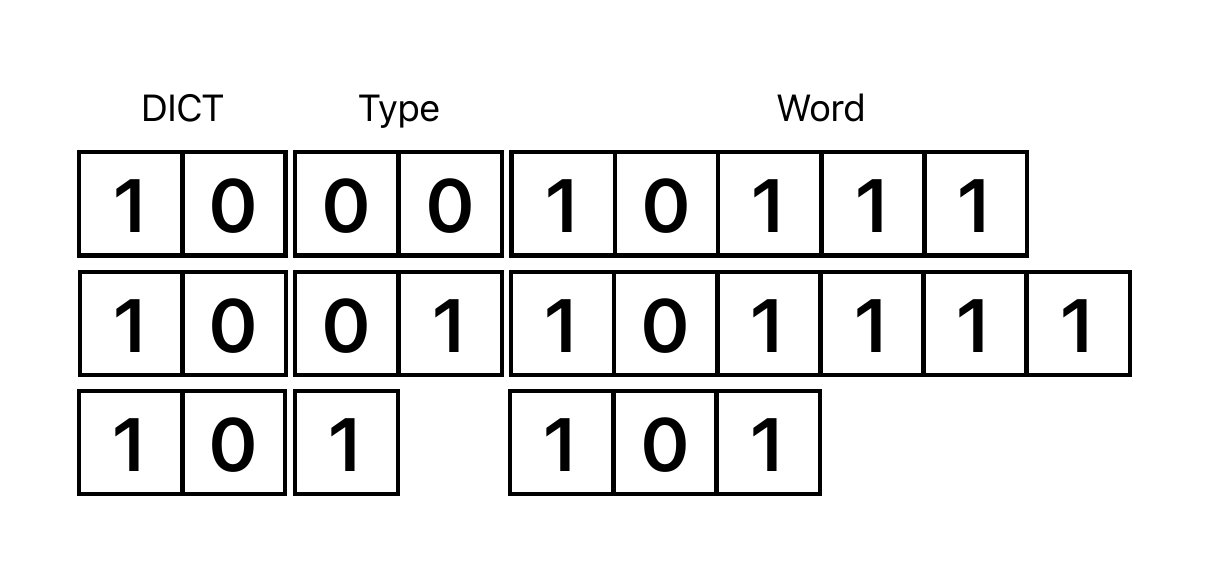}
	\end{center}
    \caption{Binary format of DICT data type}
    \label{fig:dictBinaryFormat}
\end{figure}

\end{itemize}

\subsubsection{Integer: INT16 and INT32}
\label{sub:integer}
Both INT16 and INT32 are encoded with the \textit{two's complement} \cite{patterson16} format on $16$ and $32$ bits, respectively.

\subsubsection{Real: FP16 and FP32}
\label{sub:real}
The FP16 format encodes the real number on $16$ bits using the half-precision floating-point encoding according to the IEEE Standard for Floating-Point Arithmetic (IEEE 754) \cite{IEEE754-2019}, which consists of 1 sign bit, 5 exponent bits, and 10 fraction bits.
The FP32 format encodes the real number on $32$ bits using the single-precision floating-point encoding according to the IEEE Standard for Floating-Point Arithmetic (IEEE 754) \cite{IEEE754-2019}, which consists of 1 sign bit, 8 exponent bits, and 23 fraction bits.

\subsection{QRtree header}
\label{sec:header}
After the general \textit{QRscript header}, there is optionally a \textit{QRtree header}. In order to identify the presence of this header, the first bit of the \textit{QRtreebytecode} (the one between the finish of the QRscript header and the beginning of the QRtree header) is \texttt{1}; alternatively, that bit is set to \texttt{0} to notify the absence of a QRtree header.

In case the \textit{QRtree header} is present, the following bits represent the contents of the header, which consists of a series of commands encoded on $3$ bits. The \texttt{110} code has not yet been associated with any specific command and can be used freely. This group of 3 bits is extensible following the exponential encoding provided in Appendix~\ref{app:C}.

\subsubsection{HEADER_END (000)}
The \texttt{HEADER_END} command, encoded with the \texttt{000} code, identifies the end of the QRtree header. In particular, a QRtree header is a list of commands, and since there is no indication of the number of commands present in the QRtree header, the command \texttt{HEADER_END} is used in order to identify the end of this list.

\subsubsection{INT_TYPE (001) and FLOAT_TYPE (010)}
The commands \texttt{INT_TYPE}, identified by the code \texttt{001}, and \texttt{FLOAT_TYPE}, identified by the code \texttt{010}, are used to define the storing dimensions for integer and floating point numbers, respectively.

In particular, regarding the \texttt{INT_TYPE} instruction, if the following bit is set to \texttt{0}, all the integer numbers inserted in the QRtreebytecode are to be interpreted as stored using the INT16 format, while if the following bit is set to \texttt{1}, all the integer numbers inserted in the QRtreebytecode are to be interpreted as stored using the INT32 format.

Regarding the \texttt{FLOAT_TYPE} instruction, if the following bit is set to \texttt{0}, all the real numbers inserted in the QRtreebytecode are to be interpreted as stored using the FP16 format, while if the following bit is set to \texttt{1}, all the real numbers inserted in the QRtreebytecode are to be interpreted as stored using the FP32 format.

\subsubsection{DICT_TYPES (011)}
\label{sub:DICTTYPES}
The \texttt{DICT_TYPE} command, identified by the code \texttt{011}, allows managing the activation of the \textit{global} and \textit{specific} types of dictionary for the current eQR code. In particular, the following two bits identify which dictionaries are activated by setting the options \texttt{NO_DICT_GLOBAL} to disable the \textit{global} dictionary, \texttt{DICT_GLOBAL} to enable the \textit{global} dictionary, \texttt{NO_DICT_SPEC} to disable the \textit{specific} dictionary and \texttt{DICT_SPEC} to enable the \textit{specific} dictionary. Please note that if the \texttt{DICT_TYPES} command is not present, the \texttt{DICT_GLOBAL} and \texttt{DICT_SPEC} options are enabled by default.

Table~\ref{tab:dicttypes} shows the possible configurations of the two bits following those used to identify the command (i.e., \texttt{011}) in order to activate the different combinations of options. Although the sequence of bits \texttt{11} has been reported in the table, it should never be used because the configuration with the \texttt{DICT_GLOBAL} and \texttt{DICT_SPEC} options is the default one.

\begin{table}[H]
  \caption{Bits for activating the \textit{global} and/or \textit{specific} dictionaries.}
  \label{tab:dicttypes}

  \begin{center}
    \tabcolsep=0.18cm
    \def\arraystretch{1.28}
    \begin{tabular}{c|cc}
    Bits & \textit{global} & \textit{specific}  \\
    \hline
    \texttt{00} & \texttt{NO_DICT_GLOBAL} & \texttt{NO_DICT_SPEC} \\
    \texttt{01} & \texttt{NO_DICT_GLOBAL} & \texttt{DICT_SPEC} \\
    \texttt{10} & \texttt{DICT_GLOBAL} & \texttt{NO_DICT_SPEC} \\
    \texttt{11} & \texttt{DICT_GLOBAL} & \texttt{DICT_SPEC} \\
   \end{tabular}
  \end{center}
\end{table}

\subsubsection{DICT_SPEC_TYPE (100)}
\label{sub:DICTSPECTYPE}
The \texttt{DICT_SPEC_TYPE} command, identified by the \texttt{100} code, identifies which dictionary of type \textit{specific} should be used for the current eQR code. Specifically, the following bits represent the integer index that identifies the \textit{specific} dictionary. When such dictionaries are loaded by the application, they are loaded in a certain order, and index 0 will be assigned to the first \textit{specific} dictionary loaded, index 1 to the second \textit{specific} dictionary loaded, etc. To load multiple dictionaries of type \textit{specific}, the \texttt{DICT_SPEC_TYPE} command must be repeated several times, and specifically once for each \textit{specific} dictionary loaded.

\subsubsection{DICT_LOCAL (101)}
\label{sub:DICTLOCAL}
As previously pointed out, strings are a kind of data that takes up a lot of space. Through a dictionary of type \textit{local}, it is possible to define some strings (typically those repeated several times within the eQR code) and associate them with a numeric index.

The command \texttt{DICT_LOCAL}, identified by the code \texttt{101}, allows the definition of a dictionary of type \textit{local}, whose content is saved within the eQR code itself. The first three bits after the \texttt{DICT_LOCAL} command (i.e., the bits following \texttt{101}) represent the language associated with this dictionary. These three bits follow the exponential encoding as described in Appendix~\ref{app:C}.
The \texttt{000} language is the \textit{default} language, the \texttt{001} language is language number 1, and so on. If a given dictionary is defined for multiple languages, you must repeat the \texttt{DICT_LOCAL} command as many times as the number of languages for which the dictionary is defined. The association of the language number to the specific language is defined within the application, except for the language with index \texttt{000}, which is always associated with the \textit{default} language.

After the 3 bits that identify the language, there are 4 bits (following, once again, the exponential encoding described in Appendix~\ref{app:C}) that identify the \textit{number of words} present in the dictionary. Next, there is a list of \textit{words}, which are in number equal to the \textit{number of words}. Each individual \textit{word} is identified by the encoding type (\texttt{0} for ASCII-7, or \texttt{1} for UTF-8) and the word encoded in ASCII-7 or UTF-8, respectively. In the event that a particular word belonging to a dictionary of a language other than that of the \textit{default} dictionary is encoded with the empty string (i.e., ``''), the corresponding dictionary \textit{default} word will be printed. If a word in the \textit{default} dictionary has an empty string associated with it, the empty string will be coherently printed when this word is printed.

\subsubsection{USER_DEF (110)}
The \texttt{USER_DEF} command, identified by the \texttt{110} code, has not been associated with any specific actions. It can therefore be freely used by the application developer to add new functionality to the application. It is the responsibility of the developer to maintain a correct parsing of this command in the client application that will execute the eQR code.

\subsection{Code}
\label{sec:code}
Within an eQR code, after the header due to the \textit{QRscript header} and the optional header due to the \textit{QRtree header}, there is the section called \textit{code}, which contains the language of type \textit{QRtree} that must actually be executed.

As already mentioned, this specification will detail the conversion rules from the \textit{intermediate representation} to the \textit{QRtreebytecode} (arrow 2 in Fig.~\ref{fig:generationAndExecution}). The opposite conversion, i.e. from the \textit{QRtreebytecode} to the \textit{intermediate representation} (arrow 5 in Fig.~\ref{fig:generationAndExecution}), follows exactly the application of the same rules in reverse.

From the perspective of the \textit{intermediate representation}, the \textit{code} section consists of a list of 0 or more instructions. There are 7 types of instructions, which are identified with 3 bits: \texttt{input} (\texttt{000}), \texttt{inputs} (\texttt{001}), \texttt{print} (\texttt{010}), \texttt{printex} (\texttt{011}), \texttt{goto} (\texttt{100}), \texttt{if} (\texttt{101}), and \texttt{ifc} (\texttt{110}). Code \texttt{111} has been left without a corresponding instruction to allow, through the exponential encoding described in Appendix~\ref{app:C}, to extend the instruction set of the intermediate representation and coherently of the generated QRtreebytecode.

Appendix~\ref{app:B} shows a small example of the \textit{QRtreebytecode} generated from an \textit{intermediate representation}.

\subsubsection{Direct and indirect input}
For what concerns the input instructions, there are two types:    \textit{direct} (\texttt{inputs}), and \textit{indirect} (\texttt{input}).

An input of type \textit{direct} is a text string, while an input of type \textit{indirect} is an input in which the possible choices are prefixed within a predefined set and can be defined, for example, through a set of decision buttons. The possible predetermined choices (i.e., the set of possible choices) are directly inferred by the application by analyzing the \textit{intermediate representation}. For each question, there is an additional default answer named ``Other'', which is only added if there is a chained question that follows the current one. This answer, if selected, causes the program to jump to the next chained question. Actually, if the default language used in the application is different than English, the word ``Other'' can be directly translated by the application to the target language.

\begin{figure}
    \begin{center}
	\includegraphics[width=0.4\columnwidth]{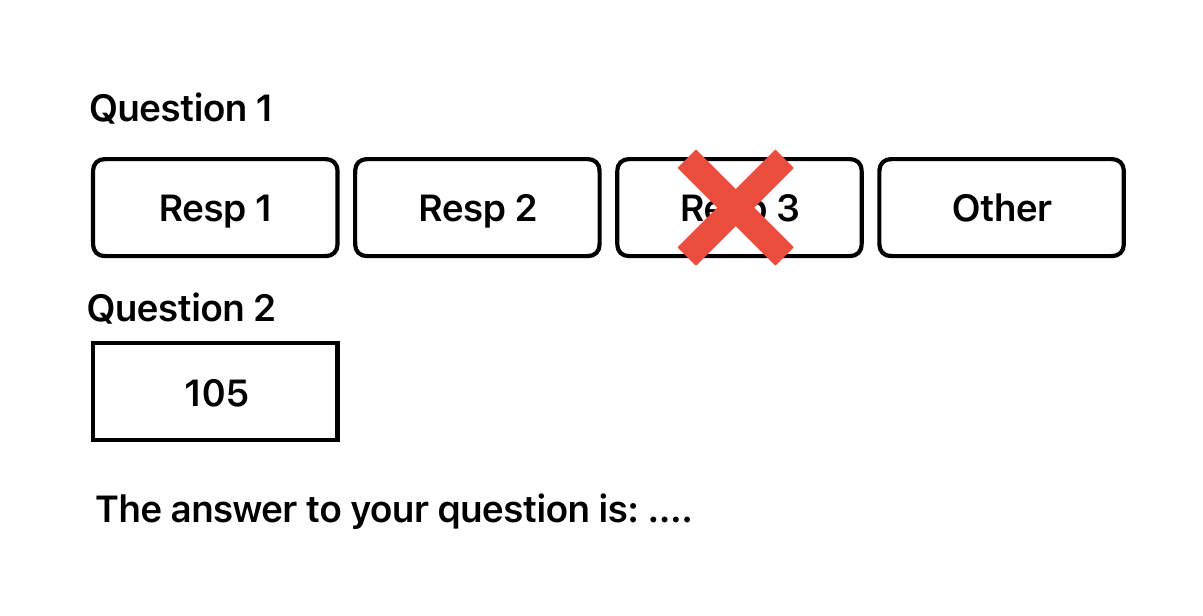}
	\end{center}
    \caption{Example of interaction between the user and the application executing the eQR code}
    \label{fig:interactionExample}
\end{figure}

As an example, the following code has been written using the \textit{intermediate representation}:
\begin{alltt}
(1)  input "Question 1"
(2)  if "Resp 1" (6)
(3)  if "Resp 2" (8)
(4)  if "Resp 3" (10)
(5)  goto (12)
(6)  # Code related to Resp 1
(7)  ...
(8)  # Code related to Resp 2
(9)  ...
(10) # Code related to Resp 3
(11) ...
(12) inputs "Question 2"
\end{alltt}
Line (1) writes the string ``Question 1'' to the screen, and an input of type \textit{indirect} is inserted (\texttt{input} command). The set of possible choices consists of \texttt{\{ ``Resp 1'', ``Resp 2'', ``Resp 3'', ``Other'' \}}. An example of a possible input interface based on decision buttons is reported in Figure~\ref{fig:interactionExample} and explained in detail in the following paragraph.

In the event that the user chooses \texttt{``Resp 1''}, for example, by clicking on the button identified by \texttt{``Resp 1''}, a jump is performed to line (6), in which the code related to \texttt{``Resp 1''} starts. Please note that this code may recursively contain other questions (i.e., a decision sub-tree). Similarly, if the user chooses \texttt{``Resp 2''}, for example, by clicking on the button identified by \texttt{``Resp 2''}, a jump is performed to the line (8), in which the code related to \texttt{Resp 2} starts. The behavior is similar even if the user chooses \texttt{``Resp 3''}. In case the user chooses the \texttt{``Other''} response, the unconditional jump instruction \texttt{goto (12)} takes the execution flow to the line (12), where there is a subsequent question, in this case, of type \textit{direct}. Since the question is direct, the user will have the possibility to enter a string that, depending on subsequent commands, can be interpreted as a string (instruction \texttt{if}) or as a number (instruction \texttt{ifc}).

\subsubsection{References}
eQR codes, but also QR codes in general, being printed on physical supports, have the undeniable advantage of being able to be combined with additional information that can be printed or placed near the eQR code itself, such as text or images. In the case the eQR code is displayed on a monitor, there is the possibility to combine it also with videos. So, in addition to a possible explanation of how the eQR code can be used, additional information associated with it, can be added and be an integral part of the program itself. For example, if a program of type \textit{QRtree} outputs very large strings, these strings (which would take up a large part of the available space within the eQR code) could be extracted from the eQR code in order to be printed near to it. This also applies to the questions that can be inserted into \texttt{input} or \texttt{inputs} instructions. For this reason, the ability for input and output instructions to print a \textit{reference} has been defined. The term \textit{reference} refers to an integer value that identifies the text or image (or any other kind of information) printed/displayed near the eQR code. For example, a very long answer is replaced with an indication to read the \textit{reference} number 1. The same reasoning applies to images: by associating an integer with an image (possibly accompanied by a textual description), the program can guide the user to follow the instructions in the image with that particular number in order to obtain information from the program itself (for example the path to reach a certain destination).
References are encoded following the exponential encoding described in Appendix~\ref{app:C}, with a starting size of $4$ bits.

All the instructions and the rules for converting from \textit{intermediate representation} to \textit{QRbytecode} are detailed in the following subsections.

\subsubsection{input (000)}
The \texttt{input} instruction is used to prompt the application end user to enter a string as an input of type \textit{indirect}.
This instruction, before requesting the insertion of the input, displays the constant \texttt{<constant>} on the screen. The instruction \texttt{input <type> <constant>} is identified by the first 3 bits equal to \texttt{000}, followed by a bit identifying the type of the constant (i.e., \texttt{<type>}), and the \texttt{<constant>} to be printed on the screen. In the case where \texttt{<type>} is \texttt{0}, the \texttt{<constant>} is to be interpreted as a \textit{string}, while in the case where \texttt{<type>} is \texttt{1}, the  \texttt{<constant>} is to be interpreted as a \textit{reference}, as detailed in Subsection~\ref{sub:string}.

\begin{figure}[H]
    \begin{center}
	\includegraphics[width=0.35\columnwidth]{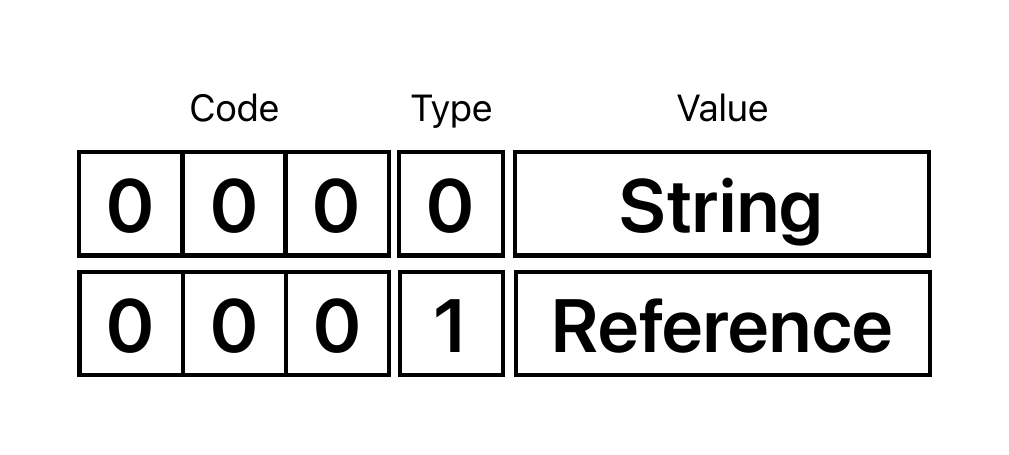}
	\end{center}
    \caption{Conversion of \texttt{input} instruction to QRtreeBytecode}
\end{figure}

\subsubsection{inputs (001)}
The \texttt{inputs} instruction is used to prompt the application end user to enter a string as an input of type \textit{direct}.
This instruction, before requesting the insertion of the input, displays the constant \texttt{<constant>} on the screen. The instruction \texttt{inputs <type> <constant>} is identified by the first 3 bits equal to \texttt{001}, followed by a bit identifying the type of the constant (i.e., \texttt{<type>}), and the \texttt{<constant>} to be printed on the screen. In the case where \texttt{<type>} is \texttt{0}, the \texttt{<constant>} is to be interpreted as a \textit{string}, while in the case where \texttt{<type>} is \texttt{1}, the  \texttt{<constant>} is to be interpreted as a \textit{reference}, as detailed in Subsection~\ref{sub:string}.

\begin{figure}[H]
    \begin{center}
	\includegraphics[width=0.35\columnwidth]{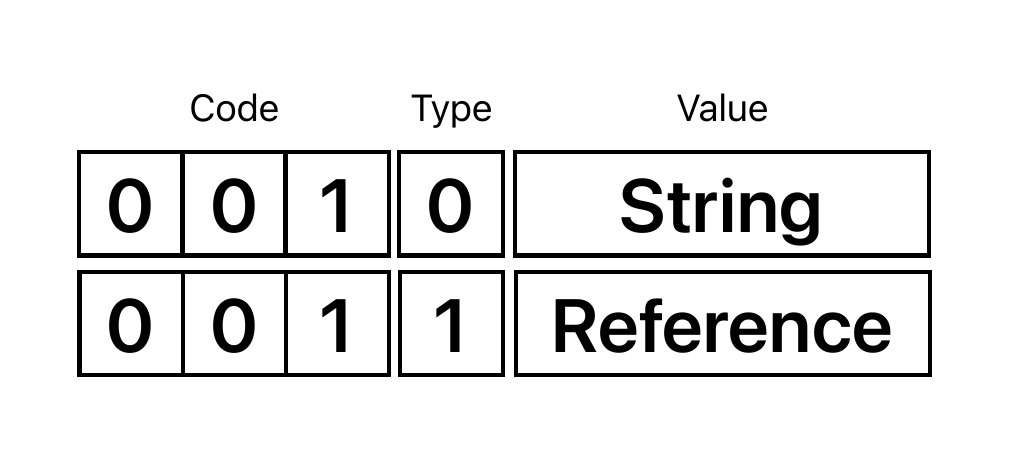}
	\end{center}
    \caption{Conversion of \texttt{inputs} instruction to QRtreeBytecode}
\end{figure}

\subsubsection{print (010)}
The \texttt{print} instruction is used to request the printing of a \texttt{<constant>} on the screen. The instruction \texttt{print <type> <constant>} is identified by the first 3 bits equal to \texttt{010}, followed by a bit identifying the type of the constant (i.e., \texttt{<type>}), and the \texttt{<constant>} to be printed on the screen. In the case where \texttt{<type>} is \texttt{0}, the \texttt{<constant>} is to be interpreted as a \textit{string}, while in the case where \texttt{<type>} is \texttt{1}, the  \texttt{<constant>} is to be interpreted as a \textit{reference}, as detailed in Subsection~\ref{sub:string}.

\begin{figure}[H]
    \begin{center}
	\includegraphics[width=0.35\columnwidth]{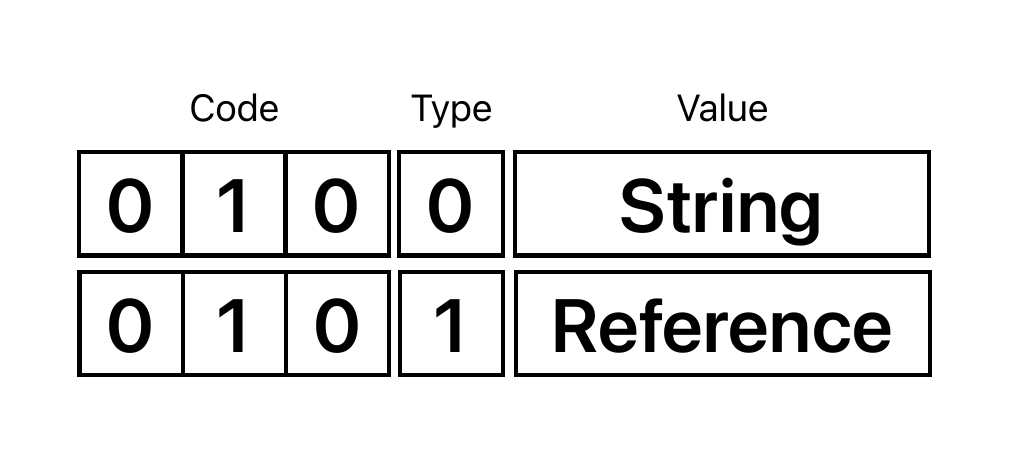}
	\end{center}
    \caption{Conversion of \texttt{print} instruction to QRtreeBytecode}
\end{figure}

\subsubsection{printex (011)}
The \texttt{printex} instruction is used to request the printing of a \texttt{<constant>} on the screen and to terminate the execution of the program. The instruction \texttt{printex <type> <constant>} is identified by the first 3 bits equal to \texttt{011}, followed by a bit identifying the type of the constant (i.e., \texttt{<type>}), and the \texttt{<constant>} to be printed on the screen. In the case where \texttt{<type>} is \texttt{0}, the \texttt{<constant>} is to be interpreted as a \textit{string}, while in the case where \texttt{<type>} is \texttt{1}, the \texttt{<constant>} is to be interpreted as a \textit{reference}, as detailed in Subsection~\ref{sub:string}. When used for printing the empty string, for example, by printing the UTF-7 character \texttt{0000011} via the instruction \texttt{011 1 00 0000011}, it causes the end of the program without printing anything on the screen. In particular, referring to the sequence of bits \texttt{011 1 00 0000011}, the first 3 bits \texttt{011} represent the \texttt{printex} instruction, the next bit \texttt{1} indicates to interpret the \texttt{<constant>} as a \textit{string}, \texttt{00} identifies the string as ASCII-7, and finally \texttt{0000011} encodes the string terminator \textit{end-of-text (ETX)} which in this context represents the empty string.

\begin{figure}[H]
    \begin{center}
	\includegraphics[width=0.35\columnwidth]{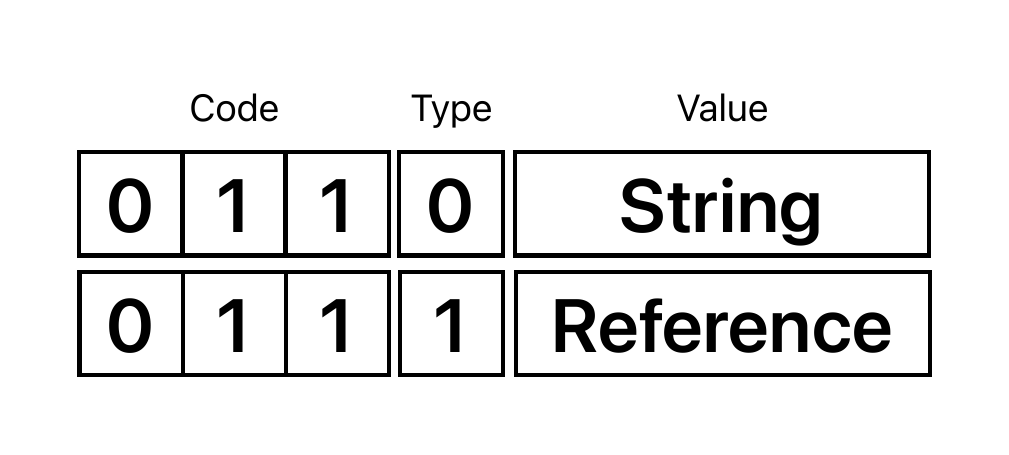}
	\end{center}
    \caption{Conversion of \texttt{printex} instruction to QRtreeBytecode}
\end{figure}

\subsubsection{goto (100)}
\label{sub:goto}
The \texttt{goto} instruction is used to perform \textit{unconditional} jumps.

The instruction \texttt{goto <relative_jump>} is identified by the first 3 bits equal to \texttt{100}, followed by a \texttt{<relative_jump>}, which represents the number of subsequent instructions for which the jump should be made. It is important to note that \texttt{<relative_jump>} only allows forward jumps. Since backward jumps are not possible, it is not possible to create cycles with \textit{QRtree}. This choice is due to the fact that, for the realization of decision trees, cycles are not of direct use. The value of \texttt{<relative_jump>} is encoded on 4 bits (following the exponential encoding described in Appendix~\ref{app:C}), and the sequence \texttt{0000} encodes a jump to the next instruction.

\begin{figure}[H]
    \begin{center}
	\includegraphics[width=0.7\columnwidth]{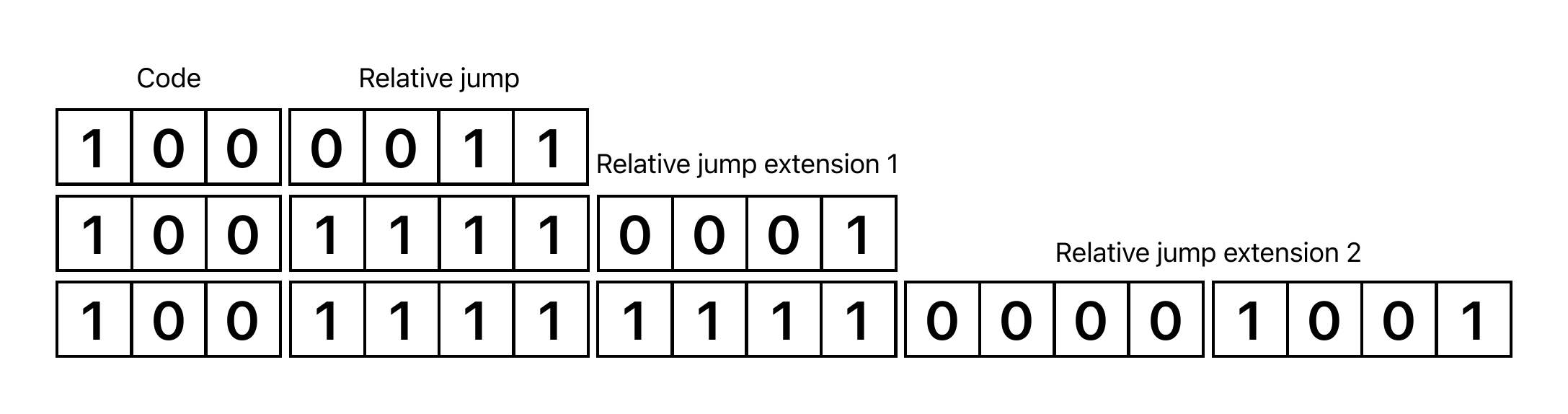}
	\end{center}
    \caption{Conversion of \texttt{goto} instruction to QRtreeBytecode}
\end{figure}

\subsubsection{if (101)}
The \texttt{if} instruction is used to perform \textit{conditional} jumps. 

The instruction \texttt{if <type> <constant> <relative_jump>} is identified by the first 3 bits equal to \texttt{101}, followed by a bit identifying the type of the constant (i.e., \texttt{<type>}), the \texttt{<constant>} (which if it is equal to the last input allows performing the jump) and by a \texttt{<relative_jump>}, which represents the number of subsequent instructions for which the jump should be made. In the case where \texttt{<type>} is \texttt{0}, the \texttt{<constant>} is to be interpreted as a \textit{string}, while in the case where \texttt{<type>} is \texttt{1}, the \texttt{<constant>} is to be interpreted as a \textit{reference}, as detailed in Subsection~\ref{sub:string}.

Each time an input instruction is executed, either \textit{indirect} (i.e., \texttt{input} instruction) or \textit{direct} (i.e., \texttt{inputs} instruction), the inserted data is saved in a variable intended to contain the last input, which will be named \texttt{tmp_input}. Whenever the content saved in the \texttt{tmp_input} variable is equal to the content saved in \texttt{<constant>}, a relative jump equal to the number of statements encoded in \texttt{<relative_jump>} is performed. If the content of \texttt{tmp_input} is different from the content saved in \texttt{constant}, the \texttt{if} instruction has no effect, and the next statement is executed. For details on how \texttt{<relative_jump>} is encoded, refer to the description of the \texttt{goto} instruction in Subsection~\ref{sub:goto}.

\begin{figure}[H]
    \begin{center}
	\includegraphics[width=0.65\columnwidth]{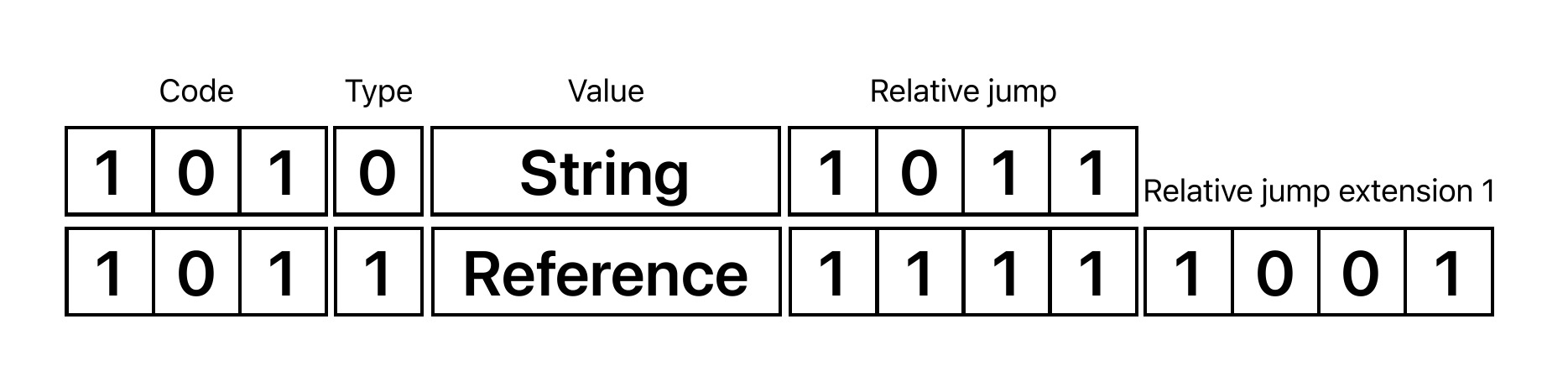}
	\end{center}
    \caption{Conversion of \texttt{if} instruction to QRtreeBytecode}
\end{figure}

\subsubsection{ifc (110)}
The instruction \texttt{ifc} is used to perform \textit{conditional} jumps after a comparison using a relational operator (i.e., \texttt{==}, \texttt{!=}, \texttt{<=}, \texttt{>=}, \texttt{<}, \texttt{>}). The comparison is executed between the value inserted via an input instruction (i.e., \texttt{input} or \texttt{inputs} instruction) and an \textit{integer} or \textit{real} number.
The instruction \texttt{ifc <rel_op> <type> <constant> <relative_jump>} is identified by the first 3 bits equal to \texttt{110}, followed by 3 bits identifying the relational operator \texttt{<rel_op>}, followed by a bit identifying the type of the constant (i.e., \texttt{<type>}), the \texttt{<constant>}, which is an \textit{integer} or \textit{real} number, and by a \texttt{<relative_jump>}, which represents the number of subsequent instructions for which the jump should be made.

In detail, the correspondence between the 3 bits of \texttt{<rel_op>} and the relational operator encoded by them are represented in the following Table~\ref{tab:rel}. The \texttt{110} and \texttt{111} bits combinations are not currently used and are left for future extensions. The \texttt{==} operator, encoded with \texttt{000}, is the equality operator, while the \texttt{!=} operator, encoded with \texttt{001}, is the inequality operator.

\begin{table}[H]
  \caption{Correspondence between \texttt{<rel_op>} and the encoded relational operator.}
  \label{tab:rel}

  \begin{center}
    \tabcolsep=0.18cm
    \def\arraystretch{1.28}
    \begin{tabular}{c|c}
    \texttt{<rel_op>} & Relational operator \\
    \hline
    \texttt{000} & \texttt{==} \\
    \texttt{001} & \texttt{!=} \\
    \texttt{010} & \texttt{<=}  \\
    \texttt{011} & \texttt{>=}  \\
    \texttt{100} & \texttt{<}  \\
    \texttt{101} & \texttt{>}  \\
   \end{tabular}
  \end{center}
\end{table}

In the case where \texttt{type} is \texttt{0}, the \texttt{<constant>} is to be interpreted as a number of type \textit{integer}, which can be encoded with the INT16 (i.e., \texttt{00}) or INT32 (i.e., \texttt{01}) notation as described in Subsection~\ref{sub:integer}). In the case where \texttt{type} is \texttt{1}, \texttt{<constant>} is to be interpreted as a number of type \textit{real}, which can be encoded with the FP16 (i.e., \texttt{10}) and FP32 (i.e., \texttt{11}) notation as described in Subsection~\ref{sub:real}. It is worth noting that depending on the setting of the header, more compact notations of the \texttt{type} are allowed.

Each time an input instruction is executed, either \textit{indirect} (i.e., \texttt{input} instruction) or \textit{direct} (i.e., \texttt{inputs} instruction), the inserted data is saved in a variable intended to contain the last input, which will be named \texttt{tmp_input}.
Whenever the comparison between \texttt{tmp_input} and the content saved in \texttt{<constant>} (using the \texttt{<rel_op>} relational operator) gives a \textit{true} result, a relative jump equal to the number of statements encoded in \texttt{<relative_jump>} is performed.
In case the comparison gives a \textit{false} result, the \texttt{ifc} instruction has no effect, and the next statement is executed. For details on how \texttt{<relative_jump>} is encoded, refer to the description of the \texttt{goto} instruction in Subsection~\ref{sub:goto}.

\begin{figure}[H]
    \begin{center}
	\includegraphics[width=0.65\columnwidth]{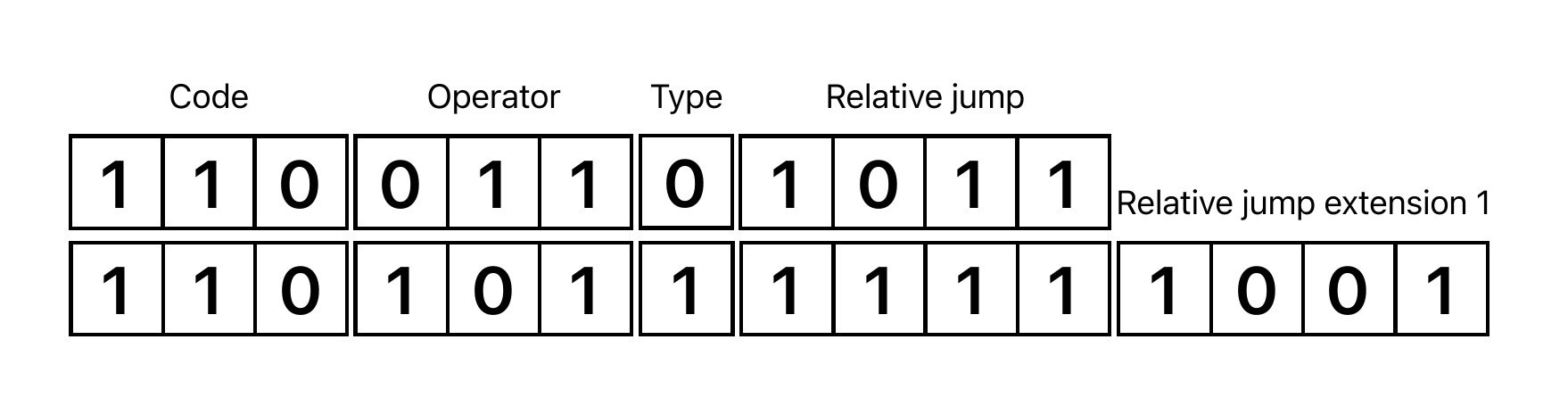}
	\end{center}
    \caption{Conversion of \texttt{ifc} instruction to QRtreeBytecode}
\end{figure}

\section{Sample implementation}
A sample implementation of this specification document is freely available under the GPL-3.0 license \cite{QRtree-soft}.

\section{Main terms and acronyms}

\begin{itemize}

    \item \textit{Continuation}: feature of the QRscript header needed to split the program over multiple eQR codes.

    \item \textit{Decision tree}: a decision tree is composed of three kinds of nodes: \textit{decision}, \textit{end}, and \textit{chance} nodes, but in the context of the QRtree dialect chance nodes have not been implemented. The decision nodes of a tree are the points where the tree branches out, and they depict the criteria that enable the user to select from various paths based on their responses. The tree's terminal nodes are the concluding points, which are utilized to produce a response to the user.

    \item \textit{Dialect}: sub-language defined through QRscript in the eQR code with implementation characteristics specific to the particular application scope (e.g., decision trees, etc.).

    \item \textit{Dialect code}: program coded within the eQR codes, which specification is reported in separate documents (e.g., in this document, the specification of the QRtree dialect is reported).

    \item \textit{eQR code}: a QR code containing executable code. It can be obtained by encoding a program within a QR Code using QRscript.

    \item \textit{eQRbytecode}: the binary representation of the executable program. It is used to create an eQR code and, after the scanning of the eQR code, to reconstruct the intermediate representation.

    \item \textit{eQRtreebytecode}: the binary representation of the part related to the QRtree dialect of the executable program. It is the last part of an eQRbytecode.
        
    \item \textit{High-level language}: a computer programming language (even visual) designed to be easy for humans to read and write. It is easier to use and more abstract than the intermediate representation. In the context of this specification document, it is a high-level programming language that can be translated into the intermediate representation of the QRtree dialect.
    
    \item \textit{Intermediate representation}: lower-level representation of the high-level language used to standardize and simplify the translation between the latter and the correspondent eQRbytecode.
    
    \item \textit{QRbytecode}: the binary representation of the executable program. It is used to create an eQR code and, after the scanning of the eQR code, to reconstruct the intermediate representation.
    It is formed by a header named \textit{QRscript header}, an optional header named \textit{QRtree header}, and by the encoding of the executable code.

    \item \textit{QRscript}: a series of rules on how to embed a programming language into an eQR code. In particular, it refers to the definition of the binary code that has to be saved inside of an eQR code and the rules used for its decodification.

    \item \textit{QRscript header}: part of an eQR code that is independent of the dialect.

    \item \textit{QRtree}: dialect mainly aimed at coding decision trees without chance nodes.

    \item \textit{QRtree code}: part of an eQR code containing the instructions of the QRtree dialect.

    \item \textit{QRtree header}: part of an eQR code containing the optional header of the QRtree dialect.

    \item \textit{QRtreebytecode}: binary representation of the part related to the QRtree dialect.    

    \item \textit{Reference}: the possibility of some instructions of the QRtree dialect to print in output a number instead of a string that represents a reference to information that is printed near the eQR code. References are used to save space by avoiding inserting long strings within the eQR code or referring to information that cannot be included in an eQR code, such as big images or videos.

    \item \textit{Security}: features of the QRscript header aimed at managing \textit{authenticity} and \textit{integrity}, and eventually the \textit{encryption} of the content of the eQR code.
    
    \item \textit{Security profile}: indicates the security mechanisms activated and exploited in the current eQR code.

    \item \textit{Three-address code}: it is an intermediate representation, and in the context of this specification document, it represents the intermediate representation of the QRtree dialect.

     \item \textit{URL}: used to permit the connection to an external uniform resource locator (URL) if an Internet or local connection is available.

\end{itemize}

\bibliographystyle{IEEEtran}
\bibliography{bibliography}

\section*{Appendix}
\appendix
\section{Definition of global and specific dictionary}
\label{app:A}

A \textit{global} or \textit{specific} dictionary can contain one or more languages. Each language must contain a language identifier on $2$ or more characters (for example, ``it'', ``en'', ``ru'', \dots) and a list of words. The order of the words is important because the index/position inside the list is the identifier used to refer to the word.

The schema for a dictionary is the following:
\begin{alltt}
"\$schema": http://json-schema.org/draft-07/schema
definitions:
  dictionary:
    type: object
    properties:
      language:
        type: string
        minLength: 2
      words:
        type: array
        items:
          type: string
oneOf:
- "\$ref": "#/definitions/dictionary"
- type: array
  items:
    "\$ref": "#/definitions/dictionary"
\end{alltt}

An example
\begin{alltt}
- language: en
  words:
    "Yes"
    "No"
    "Cancel"
    "Ok"
- language: it
  words:
    "Si"
    "No"
\end{alltt}

\section{Example of QRtreebytecode generation}
\label{app:B}


The idea behind this example is to guide the user to manage correctly a defibrillator in case it is needed. The first question asked by the program regards the heart beats that can be counted in 10 seconds, in order to get a general idea of how fast the heart is beating.

In case this number is lower than 5, then the program suggests to, first of all, call an ambulance and then proceed with the cardiac massage and the defibrillator, also giving instructions (in the case of the example, the instructions are given as references to possible images printed near the eQR code that may better explain the procedures). Then, the program asks if the patient woke up; if she/he did, it suggests waiting for the ambulance by keeping the patient awake. If she/he did not, it suggests repeating alternating the precedent procedures until the ambulance arrives.

If, instead, the number of beats in the 10 seconds is less than 10, the program suggests having the person get seated and call the ambulance only if she/he does not feel better after some time.

If the number of heart beats is more than 30, then the program notes that the person has a really high heart rate, suggesting laying the person down, making her/him breathe, and call the ambulance if she/he does not feel well after some time.

Finally, if the heart rate is not in any of the previous cases (so between 10 and 20 heart beats), the program notes that the heart rate is fine and terminates.

The high-level program used in this example is the following:

\begin{alltt}
inputs "How many heart beats can you count in 10 seconds?"
ifc <= 5:
    print "The person has a heart beat rate too low. You should call an
    ambulance."
    print "In the mean time you should try doing a cardiac massage and then
    using the defibrillator."
    print 1 # "Apply the two pads to the bare chest. One in the middle of 
    the chest and one on the left side slightly below the nipple."
    print 2 # "Don't touch the body and start the defibrillator."
    input "After a couple of charges does the patient woke up?"
    if "Yes":
        print "Great! Wait for the ambulance and keep the patient awake."
    else if "No":
        print "Keep alternating between 2-3 minutes of cardiac massage and 
        defibrillator charges until the ambulance arrives."
else ifc <= 10:
    print "The person has a slightly low heart beat. Sit them down."
    print "If they don't feel better after a couple of minutes it's better
    to call an ambulance."
else ifc >= 30:
    print "The person has a very high heart rate. Lay them down and make
    them do deep breaths."
    print "If they don't feel better after a couple of minutes it's better
    to call an ambulance."
else:
    print "The heart beat is normal."
\end{alltt}

The intermediate representation generated by the previous high-level language program through the rules specified in this specification document is the following:

\begin{alltt}
(0) inputs "How many heart beats can you count in 10 seconds?"
(1) ifc <= 5 (6)
(2) ifc <= 10 (18)
(3) ifc >= 30 (21)
(4) print "The heart beat is normal."
(5) goto (23)
(6) print "The person has a heart beat rate too low. You should call an
ambulance."
(7) print "In the mean time you should try doing a cardiac massage and then
using the defibrillator."
(8) print 1
(9) print 2
(10) input "After a couple of charges does the patient woke up?"
(11) if "Yes" (14)
(12) if "No" (16)
(13) goto (17)
(14) print "Great! Wait for the ambulance and keep the patient awake."
(15) goto (17)
(16) print "Keep alternating between 2-3 minutes of cardiac massage and
defibrillator charges until the ambulance arrives."
(17) goto (23)
(18) print "The person has a slightly low heart beat. Sit them down."
(19) print "If they don't feel better after a couple of minutes it's better
to call an ambulance."
(20) goto (23)
(21) print "The person has a very high heart rate. Lay them down and make
them do deep breaths."
(22) print "If they don't feel better after a couple of minutes it's better
to call an ambulance."
\end{alltt}

This intermediate representation is then converted into a binary format and inserted into a QR code, thus generating an eQR code.

\begin{figure}[H]
    \begin{center}
	\includegraphics[width=0.9\columnwidth]{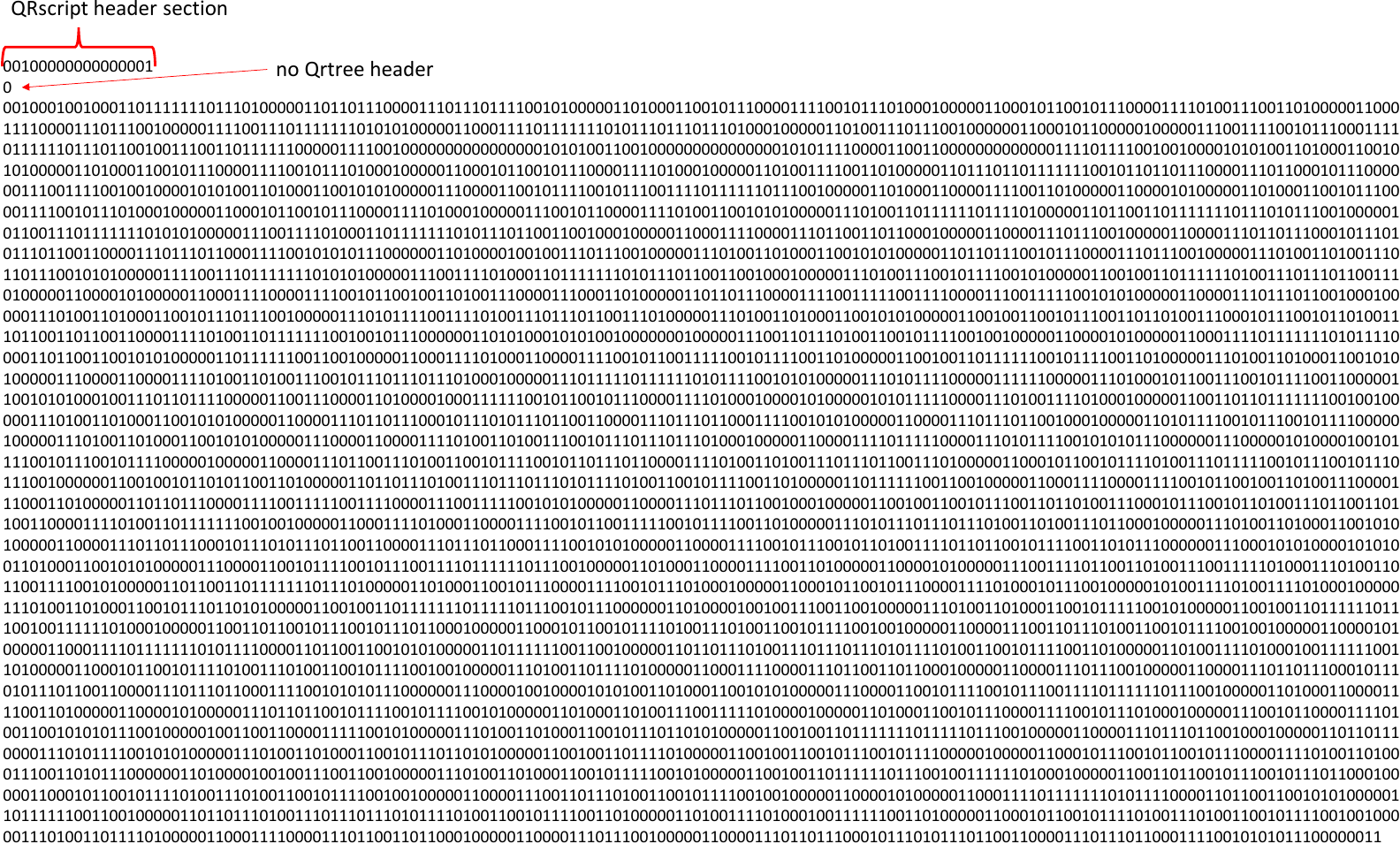}
	\end{center}
    \caption{Binary representation of the provided example}
\end{figure}

\begin{figure}[H]
    \begin{center}
	\includegraphics[width=0.5\columnwidth]{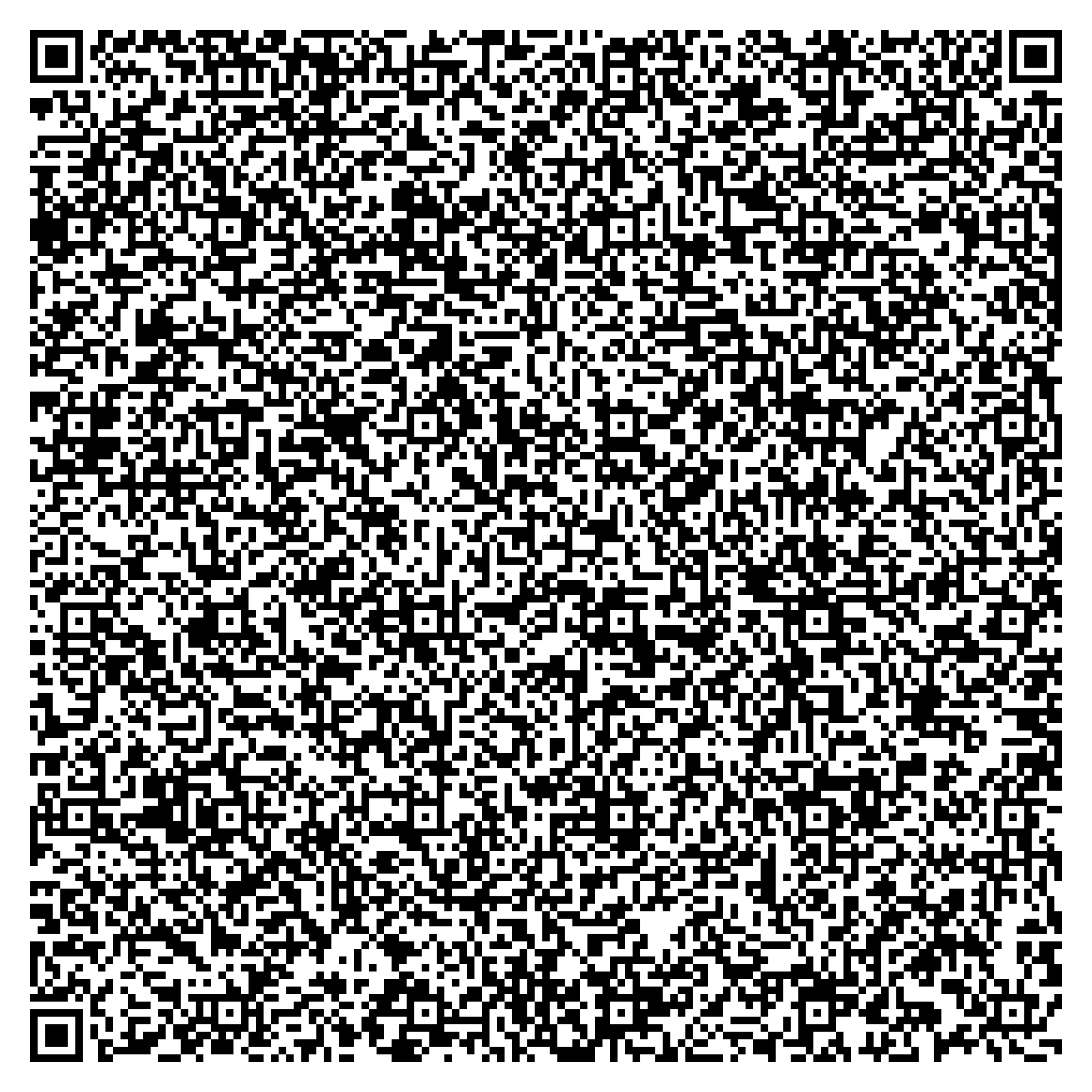}
	\end{center}
    \caption{Example of eQR code}
\end{figure}

From the conversion results between the intermediate representation and the corresponding eQRbytecode binary representation and the generated eQR code just provided, it is noticeable that the \textit{QRscript header} starts with the \textit{padding} \texttt{001}, while the \textit{continuation} is disabled by the next bit set to \texttt{0}. The \textit{security} is disabled (\texttt{0000}), and the \textit{URL} is disabled (\texttt{0}). The dialect is ``QRtree'' (\texttt{0000}), and the version is 1 (\texttt{0001}). The first bit after the \textit{QRscript header} is set to \texttt{0}, which means that the \textit{QRtree header} is also disabled.
The length of the QRbytecode is $5704$ bits, which corresponds to an occupation of the $24.1\%$ of the maximum available capacity within a QR code. Most of the space is occupied by strings and correction bits.

\section{Exponential encoding of an integer number}
\label{app:C}
The exponential encoding is used in many parts of the QRscript language to permit the encoding of integer numbers of any size, but coding at the same time small numbers with a reduced number of bits. The number of bits for the coding is doubled each time the bits used are not enough. The exponential encoding permits to have a compact notation for small numbers without limiting a priori the size of the number to be encoded.

More in detail, given an integer number without sign $X$, the exponential encoding occupies an exponentially increasing number of bits in the binary representation of the number.

Let $n$ bit be the initial number of bits used for the exponential encoding. As an example, an exponential encoding over four bits is characterized by $n=4$. If the number $X$ is greater than or equal to $(2^n - 1)$ means that the value is too big to fit in $n$ bits. In particular, the remaining value that cannot be coded in $n$ bits is $X - (2^n - 1)$. In this case, the exponential coding doubles the number of bits to $2 \cdot n$.

This process is repeated, doubling $n$ every time the number of bits is not sufficient to code number $X$ (i.e., $n_i = 2 \cdot n_{i - 1}$). The starting point ($n_0$) can be any positive integer number greater than $0$. Primarily, in this specification document, $4$ bits are used as a starting point.

Some examples of this representation can be found in Tab.~\ref{tab:expInt}.

\begin{table}[H]
    \footnotesize
    \centering
    \begin{tabular}{ c | c | c | c | c | c }
        Value & Sign & Base & $1^{st}$ extension & $2^{nd}$ extension & $3^{rd}$ extension \\
        Integer & $1$ bit & $4$ bit & $4$ bit & $8$ bit & $16$ bit \\
        \hline
        12      &   & \texttt{1100} &      &          &                  \\
        \hline
        14      &   & \texttt{1110} &      &          &                  \\
        \hline
        15      &   & \texttt{1111} & \texttt{0000} &          &                  \\
        \hline
        120     &   & \texttt{1111} & \texttt{1111} & \texttt{01011010} &                  \\
        \hline
        300     &   & \texttt{1111} & \texttt{1111} & \texttt{11111111} & \texttt{0000000000001111} \\
        \hline
    \end{tabular}
    \caption{Examples of exponential encoding of integer numbers without sign}
    \label{tab:expInt}
\end{table}

\printindex
\end{document}